\documentstyle[12pt]{article}
  \advance\oddsidemargin  by 0.5cm
  \advance\topmargin	  by -0.3in
\def\lsim{\mathrel{\rlap{\lower4pt\hbox{\hskip1pt$\sim$}}
    \raise1pt\hbox{$<$}}}	  
\def\gsim{\mathrel{\rlap{\lower4pt\hbox{\hskip1pt$\sim$}}
    \raise1pt\hbox{$>$}}}	  

\def\beq{\begin{equation}}
\def\eeq{\end{equation}}
\def\arr{\begin{eqnarray}}
\def\endarr{\end{eqnarray}}

\makeindex
\begin{document}
\hspace{12.5cm}
{\large DFTT 24/93}\\

\hspace{12.5cm}
{\large May/1993}\\
\vspace{2cm}
\begin{center}
{\Large \bf IN SEARCH FOR (QUANTUM) COLOR TRANSPARENCY}
\vspace{0.5cm}\\
{\large Boris Kopeliovich\\}
\it Dipartimento di Fisica Teorica, Universit\`a di Torino\\
\it and INFN, Sezione di Torino\\
\it Via P. Giuria 1, I-10125 Torino, Italy\\
and\\
\it Joint Institute for Nuclear Research, Dubna, Russia\\
\it e-mail: boris@bethe.npl.washington.edu\\
\vspace{2cm}
{\large \bf ABSTRACT}
\end{center}

\baselineskip  17pt

\date{ }
Color transparency (CT) is an effect of suppression of nuclear shadowing of
hard reactions, closely related to the color screening. A brief review
of theoretical development and experimental search for CT, failed and
successful, are presented. A special emphasis is  made on a
quantum-mechanical nature of CT, as opposed to a
wide spread erroneousclassical treatment of this phenomenon. The typical
predictions of the classical approach, all contradicting quantum mechanics
are:\\
{}~~-~~factorization of cross section of hard reactions on a nucleus;\\
{}~~-~~"nuclear transparency", a normalized ratio of nuclear to nucleon
cross sections, cannot exceed one;\\
{}~~-~~the larger is a radius of a hadron, the stronger it attenuates in
a nucleus;\\
{}~~-~~the higher is the energy of hadrons participating in a hard reaction,
the less is the nuclear attenuation;\\
{}~~-~~due to CT hard processes provide a better opportunity
to study Fermi-momentum distribution, than soft reactions; etc.
\vspace{2.5cm}
\begin{center}
{\sl Talk presented at XXVIII Rencontres de Moriond\\
March 1993}
\end{center}

\newpage

\baselineskip  22pt

\section{Quantum approach versus vulgar
(classical) treatment of Color Transparency. }

Nuclei are unique analyzers of the space-time evolution of strongly
interacting hadronic system at early stage of their development of
about a few fermi's.  Specifically, a nucleus can play a role of a
detector of the size of the ejectile (sometimes of the projectile as
well) emerging
from a reaction on a bound nucleon.  Of special interest are the
exclusive reactions, where any soft inelastic final (initial) state
interaction looks like an absorption of the hadron, which we are
tracing on.  The smaller is the size of the wave packet, the weaker it
interacts, the more transparent is the nuclear matter.	It is a direct
consequence of color screening:  a colorless object can interact
only due to a transverse distribution of hidden color.	The cross
section vanishes as a square of the color dipole momentum of the
state.  Another
important condition is a sufficiently high energy of the ejectile to
"freeze" its transverse size while it is passing through the nucleus.
We will be back to this question later.

The phenomenon of suppression of nuclear shadowing in some reactions
due to the color screening
is called color transparency (CT). It was first demonstrated in diffractive
processes \cite{ZKL,BBGG} and suggested in
quasielastic scattering \cite{M,B}.

One of the goals of the present talk is to give a brief review of
theoretical development and results of experimental search for CT.
The another is to emphasize the importance of quantum-mechanical
approach to CT, as distinct from a vulgar treatment of this
phenomenon, which is unfortunately wide spread.  The latter is
probably because human intuition prefers a simplification towards the
classical understanding.  We are discussing it below, but here there
are a few typical examples of misuse
of such simplifications (one can find more in \cite{FS}):

\begin{itemize}
\item
{\sf Factorization}. {\sl It is assumed that the cross section,
$\sigma_A$, of a hard process on a nucleus factorizes to the cross
section on a bound nucleon, $\sigma_N$, and a survival probability of
participating hadrons to traverse the nucleus without interaction,
\beq
Tr=\frac{\sigma_A}{A\;\sigma_N}
\label{aa}
\eeq
The quantity Tr is usually called nuclear transparency.}

A quantum-mechanical interference, however, strongly violates the
factorization.	We give below a few explicit examples
\cite{KZpr,KNNZ,JK}, when the
transparency (\ref{aa}) exceeds unity, what would be impossible,
if $Tr$ indeed were a transparency.

\item
{\sf Space-time evolution of ejectile}.  {\sl Classical approach
operates with fixed, average sizes of an initial state, produced in a
hard process, and a final hadron.  It is assumed that quarks propagate
along fixed trajectories with separation increasing as a linear or
square root function of time.  The latter is called sometimes "quantum
expansion" or "quantum diffusion".  Despite the fancy use of the word
"quantum", this approximation misses all known quantum-mechanical effects.
It predicts:  i)that nuclear transparency (\ref{aa}) is always less
that unity; ii) the larger is the hadronic radius, the stronger is the
nuclear attenuation; iii) nuclear attenuation
decreases with energy, since the expansion slows down due to Lorentz
time delay; etc.}

These expectations based on the classical treatment of the evolution,
fail in many cases, if one compare them  with results of correct
quantum mechanical
calculations.  One should use wave functions of initial and final
states, rather than average radiuses, and sum over different quark
trajectories \cite{KZpl,KZpr,KNNZ}

\item
{\sf Color filtering.} {\sl This means that large-separation
components of a wave packet propagating in a nuclear matter are
filtered out due to a stronger attenuation. The classical approach is
principally unable to incorporate this effect, because it ignores
the distribution over the transverse separation in the wave packet.}

Properly taken into account, the color filtering leads to salient
predictions. It makes a nuclear matter much more transparent
\cite{ZKL}, and increases transverse momenta of particles produced in
diffractive dissociation \cite{BBGG} The filtering changes the form
of the wave packet, what results in a nuclear antishadowing in
some channels \cite{KZpr,KNNZ,Kolen'ka}.

\item
{\sf Fermi motion in A(e,e'p)A'.} {\sl There is a wide spread opinion,
that quasielastic electron scattering at high $Q^2$ is a precise tool
for study Fermi
distribution in nuclei.  It is based upon the classical treatment of
CT and of the evolution as well. The idea is, that the final state interaction
vanishes at high $Q^2$, and the recoil proton carries an undistorted
information about its initial Fermi momentum.}

A quantum-mechanical consideration of CT \cite{JK} in hadronic basis
shows that in this specific reaction CT is possible only due to the Fermi
motion.  Different components of the nuclear wave function add up to
create a small-size wave packet, eliminating a certainty in an
initial Fermi momentum of the proton.  Besides, this wave packet
producing a proton in the final state, transfers to the nuclear matter
during passing it
a negative longitudinal momentum of uncertain amount.  The
same concerns the wide-angle quasielastic proton scattering.

\end{itemize}

\section{Space-time evolution}
The problem of evolution of a wave packet in nuclear environment
is of great importance, and probably is the most complicate one.
There are known a few approaches. The first one, exploring the
connection of the hadronic basis with
eigenstates of interaction, was suggested in 1980 \cite{KL}.
It was also noticed  in the first paper on CT
\cite{ZKL}, that CT is a particular case of Gribov's inelastic
corrections \cite{Gribov}. Later the connection with the hadronic basis
was explored also in \cite{JM1,JM2}.

If one decomposes an ejectile wave packet over the complete set of
hadronic states, one should sum all over the amplitudes of hard production
of these states, including all possible diagonal and off
diagonal diffractive rescatterings in the
nucleus.  It is very difficult
problem.  An effective approach to this difficult problem, an
approximation of diffractive matrix, was suggested recently
in \cite{NO}.

In the hadronic representation the violation of the factorization is
quite natural: one should compare a hard production of a proton on a free
proton  target, with production of different excited states on a bound
proton.

It is possible to study the evolution in the quark representation as well,
but one
has to take into account propagation of the quarks over all possible
trajectories, weighted with appropriate factors. It was done in
\cite{KZpl,KZpr,KNNZ} using the path integral technique.

\section{Diffraction on nuclei}

{\sf Nonexponential attenuation}. The effect of high nuclear
transparency originating from the color screening was
first claimed  in \cite{ZKL}. The amplitude of probability of passing
a nucleus of thickness $T$ reads,
\beq
F= \langle  f|e^{-\frac{1}{2}\;\sigma(\rho^2)\; T}|i\rangle_{\rho}
\label{3a}
\eeq
Here $|i\rangle$ and $|f\rangle$ are initial and final state wave
functions. The interaction cross section of $q\bar q$
pair with transverse separation $\rho$ behaves like $\sigma(\rho)\propto
\rho^2$ at $\rho\rightarrow 0$ due to the color screening.

In the limit $T\gg 1\;fm^{-2}$ expression (\ref{3a}) gives $F\propto
1/T$. This nonexponential attenuation results from presence in the
wave packet a penetrating component with small $\rho$. The same
component, filtered out by the nucleus, provides a broadening of
transverse momenta of hadrons produced in diffractive dissociation
\cite{BBGG}.

These manifestations of CT are completely lost in the classical
approach.
\medskip

{\sf Diffractive virtual photoproduction of vector mesons on nuclei}.
This process is a perfect laboratory for study of CT.
The qualitative space-time pattern of it was discussed in \cite{BM},
and a detailed analyses was undertaken in \cite{KZpr,OB,KNNZ}. The main
observations are:

Nuclear antishadowing, $Tr>1$, of production of radial
excitations, $\Psi'$ and $\rho'$ at small $Q^2$ \cite{KZpr,KNNZ}.
This is a direct consequence of the color filtering.  Indeed, these
states have a small overlap with a quark component of a photon due to
a node in the wave functions $\Phi_{V'}(\rho)$. The nuclear filtering
squeezes the passing wave packet, and can substantially increase the
overlap with $\Phi_{V'}(\rho)$.

$Q^2$-dependence of nuclear transparency for $\Psi'$ and $\rho'$ production
is shown in fig.1 \cite{KNNZ}. The antishadowing at small $Q^2$ changes to
an universal approach from below to $Tr=1$ at high $Q^2$.

It was predicted in \cite{KZpr}, that, contrary to the naive
expectation, the nuclear transparency in diffractive photoproduction
of vector mesons decreases at high energies. This is a consequence of
the growth of the coherence length, resulting in a longer path inside the
nucleus covered by the quark fluctuation of the photon. This
prediction was confirmed by measurements of the NMC collaboration
in a good agreement with calculated energy dependence
\cite{OB}.

The diffractive photoproduction of vector mesons provides  unique
information about their wave
functions.  Indeed, varying $Q^2$, one scans the wave function of the final
meson, changing the range of impact parameter $\rho$, where the wave
function of the $q\bar q$ wave packet and $\Phi_{V'}(\rho)$ overlap
\cite{KNNZ}.

\section{Quasielastic scattering}
{\sf Electron scattering A(e,e'p)A'}. As distinct from the diffraction,
it is not so easy to produce in this process a small-size wave packet,
consisted of many
states, $p, p^*, p^{**}...$.  It is impossible at all on a free
proton target, because the mass of the ejectile is strongly correlated with
value of Bjorken variable, $x_B=Q^2/2m_p\nu$, fixed by the electron
momenta.  One knows, however, that in quantum mechanics a detector
affects the result of measurement.  It is just the case:  putting a
detector of size of the ejectile (rescattering on other nucleons)
close to the target proton, one cannot more consider the latter as
being at rest (Fermi motion), due to the uncertainty principle.
(Fermi motion).  As a result the mass of
ejectile acquires an uncertainty too, and a wave packet of a definite
size can be produced.  However the mass spectrum of is
restricted by the available Fermi momenta, and depends on the value of
$x_B$.	Varying the latter, one changes the amount of CT: $Tr$ increases
at $x_B<1$ (positive missed momenta) and
even exceeds one.  On the contrary, $Tr$ falls down at $x_B>1$
approaching the expectation of the Glauber approximation.  Results of
calculations \cite{JK} in a simple two channel model, with $m=m_p,
m^*=1.6~GeV$, are shown in fig.2 vs $Q^2=7, 15, 30 GeV^2$. We predict
very weak effect around $x_B=1$ at $Q^2=7~GeV^2$, in agreement with
the results of recent measurements at SLAC.
The effect might be even overestimated
by this model. It was argued in \cite{NO}, using  more developed model,
that the initial size of the produced wave packet only slowly decreases
with $Q^2$.

Note that due to this uncontrolled Fermi bias of $Tr(x_B)$, CT doesn't
help, but spoils any opportunity to measure the large momentum
tail of Fermi distribution, even on light nuclei. It is
better to do at small $Q^2$. The same is true for the process which
follows.

\medskip

{\sf Wide angle A(p,2p)A' scattering.} The Fermi bias of nuclear
transparency might be closely relevant to the puzzling results of
search for CT in Brookhaven experiment \cite{BNL}: nuclear transparency
unexpectedly falls  down the value
corresponding to Glauber approximation, at incident momenta above
$10~GeV/c$. The measurements were performed
at three beam momenta, 6, 10 and 12~GeV/c, and distributed over missed
("Fermi") momenta. The effect of Fermi bias just causes the
decreasing dependence of $Tr$ on the c.m. total energy, which was
observed at beam momentum 12~GeV/c. Numerical
estimations with the same simple model are compared with the data
\cite{BNL} in fig.3. The interval of masses $m^*=1.5-1.8~GeV$ was
used. At lower beam momentum the effect is weaker due to a faster
evolution.  The calculations essentially underestimate only one point
in fig.3b. Otherwise the data do not contradict the model, though the latter
is oversimplified.

\section{Observed signals of CT}

{\sf Quasifree charge-exchange scattering}. In quasielastic scattering
presence of Landshoff-type graphs suppresses CT signal up to very high
$Q^2$, when  Sudakov form factor becomes important. In the reaction of
charge-exchange scattering the Regge poles are known to dominate at low
transferred momenta $q^2$, what provides a formfactor-type vertex. At
higher values of $q^2$ Regge cuts, containing Landshoff-type graphs,
become important. Destructive interference pole-cut manifests as a
minimum in $q^2$-dependence of differential cross section. For
instance in reaction $\pi^-p\rightarrow\pi^0n$ it occurs at
$q^2\approx 0.6~GeV^2$. Hence at $q^2<0.5~GeV^2$ one can believe in
the pole dominance.

Without Landshoff graphs the hadron formfactor squeezes the hadron
to a small size even at quite low $q^2$. Indeed the form factor
provides a size $\rho^2\sim 1/q^2$, as compared with average hadronic
dimension $\rho^2\sim m_{\pi}^2$. So the absorption cross section of
such a wave packet is suppressed by a factor of the order of
$m_{\pi}^2/q^2$, what is enough to make a nucleus transparent already
at $q^2>0.2~GeV^2$. So we have a gap $0.2<q^2<0.5~GeV^2$, where a
strong CT effect could be expected.

Result of calculations \cite{KZyaf,KZpl} are compared with data
\cite{data} in fig.4  Multiple elastic rescatterings were included,
because the recoil neutron escaped detection.  Calculations were
performed in the standard Glauber approximation and including CT effect.
One can see that the data strongly support the latter variant.
\medskip\\
Note that strong CT effect predicted and observed in Regge-exchange
amplitude, should result in a nuclear enhancement of polarization
in a quasielastic scattering. Indeed, at high energies polarization is
suppressed by smallness of a spin-flip amplitudes, dominated by
Regge-exchange. On a nucleus the latter is enhanced, but the Pomeron,
non-flip part not. So the ratio of nuclear to nucleon polarizations
is rising function of $Q^2$. On heavy nuclei this ratio exceeds factor
of 2 at $Q^2\approx 1~GeV^2$ \cite{KZyaf}
\medskip

{\sf Inclusive hadron production in deep inelastic scattering.} CT
effects are not restricted only by exclusive processes. They might be
important also in inclusive reactions, at least at a kinematic border,
towards the exclusive limit.

High nuclear transparency, $Tr\approx 1$ was observed by EM
collaboration \cite{EMC} at high energies. Calculations
\cite{KN}, taking into
account a formation zone of particle production and CT effect nicely
agree with the data, shown in fig.5 One can see that without CT
effects theory substantially underestimates the data.

\section{Conclusion}

This short talk was aimed to the emphasis of the quantum
mechanical nature of the CT phenomenon as opposed to the wide
spread classical interpretation. One can find more about this
in reviews \cite{Boren'ka,Kolen'ka}
\medskip\\

{\sf Acknowledgements}.
The author has benifitted from usefull comments by N.N. Nikolaev.
He thanks for hospitality LPTHE of
Centre de Orsay, where this talk was prepared, and Department of
Theoretical Physics of Torino University, where this report was written.

\vspace{1cm}
\baselineskip 20pt
\begin{center}
{\Large Figure captions}
\end{center}
\vspace{0.3cm}

{\bf Fig.1} Nuclear transparency for $\Psi'$ and $\rho'$ electroproduction
on $Fe$ as function of $Q^2$.
\vspace{0.3cm}

{\bf Fig.2} Nuclear transparency in $(e,e'p)$ reaction on Fe as function
of $x_B$. Curves, long-dashed, solid and short doshed, correspond to
values of $Q^2=7, 15$ and $30~GeV^2$ respectively.
\vspace{0.3cm}

{\bf Fig.3} Comparison with data \cite{BNL} at beam momenta 6 (a), 10 (b)
and 12 (c)
GeV/c. At each beam momentum the points are distributed over the missed
momentum as it is explained in \cite{BNL}. Calculations are perfermed
in the simplest two-channel model with $m^*=1.5~GeV$ (dashed curve), and
$m^*=1.8~GeV$ (solid curve).
\vspace{0.3cm}

{\bf Fig.4} Nuclear transparency of reaction $\pi^- C^{12}\rightarrow
\pi^0 X$ as function of the momentum transfer squared $q^2$.
Data at $40~GeV/c$ are from \cite{data}. The dashed curve is the
result of Glauber approximation. The solid curve incorporates with CT.
\vspace{0.3cm}

{\bf Fig.5} Nuclear transparency in the inclusive electroproduction
of hadrons, $eA\rightarrow e'hX$, at $\langle\nu\rangle =75~GeV$, as
function of $Q^2$. The dashed and dotted curves are the results of
calculations, taking into account only the effect of formation length, or
CT. The solid curve incorporates both.
\end{document}